
\input harvmac.tex

\Title{ \vbox{\baselineskip12pt\hbox{ YCTP-P16-93}\hbox{hep-th/9307085}}}
{\vbox{\centerline{ Comment on Two Dimensional $O(N)$ and $Sp(N)$}
 \vskip2pt\centerline{ Yang Mills Theories as String Theories} }}
\centerline{ Sanjaye Ramgoolam }
\centerline { Department of Physics}
\centerline { Yale University, New Haven CT 06511-8167}

\vskip .3in
  We write down  all orders large $N$ expansions for the dimensions
of irreducible representations of $O(N)$ and $Sp(N)$. We interpret
all the terms in these expansions as symmetry factors for singular
worldsheet configurations, involving collapsed crosscaps and tubes.
 We use
it to complete the interpretation of two dimensional Yang Mills
Theories  with these gauge groups, on
arbitrary two dimensional manifolds, in terms of a String Theory of maps
of the type considered by Gross and Taylor. We point out some intriguing
similarities to the case of $U(N)$ and discuss their implications.

\Date{July 1993}

\newsec{Introduction}
Gross and Taylor \ref\GrTa {D. Gross, `Two Dimensional QCD
 as a String Theory
,' PUPT-1356, LBL-33415, Hepth-9212149; D. Gross,
W. Taylor, `Two-dimensional QCD is a String Theory,' PUPT-1396, LBL-33458,
Hepth-93011068; D. Gross, W. Taylor, `Twists and Loops in the String
Theory of Two Dimensional QCD,' PUPT-1382, LBL-33767, Hepth-9303076.  },
 and Minahan \ref\Minahan{J. Minahan, ` Summing Over Inequivalent Maps
in the String
Theory of Two Dimensional QCD,' UVA-HET-92-10, Hepth-93011068.}
  have established
that the partition function,
 of two-dimensional $U(N)$ or $SU(N)$
Yang Mills theory can be expanded in $1/N$ to give
terms that all have
a geometrical interpretation in terms of maps of a
string worldsheet to
the target space. The partition function for an
 orientable closed manifold
of genus $G$ is \ref\Mig{A. Migdal, Zh. Eskp. Teor. Fiz. 69, 810
(1975)(Sov. Phys. JETP. 42 413)}\ref\Rus{B. Rusakov, Mod. Phys. Lett. A5,
693 (1990).}
\eqnn\pfun $$\eqalignno{
   Z_{\cal M} &= \int [DA^{\mu}]exp[{{-1}\over {4{\tilde g}^2}}
\int_{\cal
M} d^2x \sqrt {g} Tr F_{\mu\nu}F^{\mu\nu}] \cr
 &= \sum_{R}(\dim R)^{2-2G}
e^{-{{\lambda A}\over {2N}} C_2(R)},&\pfun\cr }$$
where $\lambda = {\tilde g}^2N$.

 Our purpose is to investigate if a similar picture
holds for $O(N)$ and $Sp(N)$ Yang Mills with the hope of
 learning more about the class
of string theories related to two dimensional Yang Mills theories.
The investigation of the stringy interpretation of $O(N)$ and $Sp(N)$
 Yang Mills
theories has been carried out in \ref\NRS{S. G. Naculich, H. A. Riggs,
and H. J. Schnitzer, `Two-dimensional Yang-Mills Theories are String
Theories,' BRX-TH-346, JHU-TIPAC-930015, Hepth/9305097'.},
 for closed Riemann surfaces (orientable or
nonorientable), for the terms coming from the casimir and the leading two
terms of the $1/N$ expansion of  $\dim R$, which suffices for
a complete description of the theory on closed manifolds of Euler character
 zero. In this paper
 we write the full
$1/N$ expansion for the dimension and identify the $\Omega$ point for
$O(N)$ and $Sp(N)$, interpreting all the coefficients
 that appear in the $\Omega$ point
in terms of a localised set of singularities.
We also identify the appropriate observables which allow a stringy
interpretation of the theory on manifolds with boundary. We will point out
some very intriguing similarities between the $\Omega$ points of $O(N)$,
$Sp(N)$ and
the omega point for `composite' representations of $U(N)$. We will refer
frequently to the `chiral' omega point ($\Omega_n$)
 and the `coupled' omega point ($\Omega_{n,\tilde n}$) of \GrTa . They
are defined in terms of the dimensions of $U(N)$ reps as
 \eqn\chir { \dim R = {{N^n}\over {n!}} \chi_{\scriptscriptstyle R}
(\Omega_n),}
where R is the representation associated to a Young tableau with $n$ boxes,
and
 \eqn\coup{ \dim (R{\overline S}) = {{N^{n+\tilde n}}\over {n!{\tilde n}!}}
\chi_{R {\overline S}} (\Omega_{n \tilde n}) }
where $R$ has $n$ boxes and $S$ has $\tilde n$ boxes.  $(R\overline S)$ is
the irreducible representation of largest dimension in the  tensor product
of $R$ with the complex conjugate of $S$.

We will discuss in detail in the sections 2-4 the case of $O(N)$ and
in section 5 we describe the small modification for $Sp(N)$.
In section 2 we quote a character formula for irreducible tensor
representations of $O(N)$ and convert
it to a form appropriate for our use. We specialise it to give
the expansion of the dimension. We will not consider the spinor
representations in detail in this section.
They have casimirs of order $N^2$ \NRS\ and do not contribute to the large
$N$ large $A$ asymptotics but may have nonperturbative effects.
In section $3$, we discuss the
geometrical interpretation of the expansion. Some of the terms have a
familiar interpretation similar to the case of the chiral expansion for
 $U(N)$. In addition there is a set of permutations which we identify as
describing multiply branched sheets equipped with a collapsed crosscap
or with an infinitesimal tube connecting two cycles. To prove that the
coefficients in the expansion are really the symmetry factors for the
singular maps described above,
we derive, in section $4$, a simple formula for a certain sum of
characters of $S_n$, called $X_{\sigma_1}$, which appears in
the character formula
for $O(N)$.  We comment on the formula for the dimensions of the spinor
representations. In section 5, we write the large $N$ expansion for
$Sp(N)$  and show that the geometrical interpretation carries over with
one modification, the `collapsed crosscaps on branch points' do not come
with a minus sign as in the case of $O(N)$.

\newsec{Formula for characters and dimensions of $O(N)$ representations }

The following formula gives the large $N$ expansion for the dimension of a
tensor representation of $O(N)$ associated with a Young tableau with $n$
boxes.
\eqnn\dimens $$\eqalignno{ \dim R
&= \sum_{\sigma \in S_{n}}
{ {\chi_{\scriptscriptstyle R}(\sigma)}\over {n!} }
\sum_{ {\scriptstyle n_1 \ge 0} }
\sum_{ {\scriptstyle T_{\sigma_1}} \atop
        {\scriptstyle {T_{\sigma_1} T_{\sigma_2} = T_{\sigma}}}
     }
N^{K_{\sigma_2}} X_{\sigma_1} {{n!} \over {n_2!}}
{{|T_{\sigma_1}||T_{\sigma_2}|}\over {|T_{\sigma}|} }
\cr
&= {{N^n}\over {n!}} \chi_{\scriptscriptstyle R}(\Omega_n) .&\dimens\cr}$$
For each $n_1$ we are summing over conjugacy classes $T_{\sigma_1}$ of
$S_{n_1}$, with $|T_{\sigma_1}|$ being the order of the conjugacy class
and $\sigma_1$ a representative of the conjugacy class. The
condition $T_{\sigma_1}T_{\sigma_2}=T_{\sigma}$ means that the cycles of
$\sigma$ can be separated into two sets one of which characterises a
conjugacy class in $S_{n_1}$ and the other a
conjugacy class in $S_{n_2}(n_2=n-n_1)$.
$X_{\sigma_1}$ is defined by
\eqn\Xsig { X_{\sigma_1} = (-1)^{n_1/2} \sum_{R \in Y_{n_1}^*}
 {{\chi_{\scriptscriptstyle R}
(\sigma_1)} \over {n_1!} } ,}
where $Y_n^*$ is a subset of the Young tableaux with $n$ (even) boxes,
which is described in the appendix.
The expression for $\Omega_n$ is
\eqnn
\Omegapt$$\eqalignno{ \Omega_n &= \sum_{\sigma \in S_n}\sigma
 \sum_{ {\scriptstyle n_1 \ge 0}}
\sum_{ {\scriptstyle T_{\sigma_1}} \atop
        {\scriptstyle {T_{\sigma_1} T_{\sigma_2} = T_{\sigma}}}
     }
N^{-n_1}X_{\sigma_1}N^{K_{\sigma_2}-n_2}{{n!}\over {n_2!}}
{{|T_{\sigma_1}| |T_{\sigma_2}|}\over {|T_{\sigma}|}}\cr
                     &= (1 + O(1/N) ) .&\Omegapt\cr}$$
The leading term in \dimens\ is obtained, when $n_1$ is zero, and
$\sigma_2$ is the identity
 element and
the next term when it is an element belonging to the conjugacy class
 characterised by $1$ cycle of
length $2$ and all other cycles of length $1$ (the class $T_2$ in the
notation of \GrTa) . These are easily seen to agree with the terms as
computed in \NRS\ from a
hook length type formula.

 We now prove equation \dimens .
Littlewood \ref\lit{D. E. Littlewood, `The Theory of Group Characters and
Matrix Representations of Groups,' Oxford, The Clarendon Press, 1940.}
(page 240) gives the
following formula for the
 characters of $O(N)$
 \eqn\charform { \chi_{\scriptscriptstyle R} (U) = \{R\} +
\sum_{n_1\ge 2}\sum_{R_1\in Y_{n_1}^*}
\sum_{\scriptstyle R_2 \in Y_{n_2}\atop\scriptstyle n_1+n_2=n}
(-1)^{n_1/2} g(R_1, R_2; R)
\{R_2\}.  }
$R$ is a Young tableau having $n$ boxes.
$R_1$ runs over a certain subset of the Young tableaux with
$n_1$ (even) boxes we will call $Y_{n_1}^{*}$ and describe
 in detail in
Appendix A. $g(R_1,
R_2; R)$ are the Littlewood-Richardson coefficients. $R_2$ runs
 over all
tableaux with $n_2$ boxes. $\{R\}$ and $\{R_2\}$
are symmetric functions in the $N$ eigenvalues
of the $O(N)$ matrix $U$.
 The  symmetric
function $\{R\}$ can be
written in terms of characters of $S_n$ :
\eqn\expS{ \{R\} = \sum_{\sigma\in S_n}
 {{\chi_{\scriptscriptstyle R}(\sigma)}\over {n!}}
 P_{\sigma}. }
$P_{\sigma}$ is a power sum symmetric polynomial,
\eqn\psig{ P_{\sigma}(x_1,x_2, \cdots) = \prod_{i}
(x_1^{r_i}+x_2^{r_i}+\cdots )^{n_i} ,}
where $i$ runs over the cycle lengths ($r_i$), which have
 multiplicity $n_i$,
 in the cycle
decomposition of $\sigma$. If there are $N$
arguments $x_1,x_2,\cdots, x_N$
which are the $N$ eigenvalues
of the matrix $U$, this is the $Y_{\sigma}(U)$ of Gross-Taylor
\GrTa .

  We will rewrite the character formula in a form appropriate for a
geometric interpretation a la Gross-Taylor. Let us
 find $X_{\sigma_1}$ such
that
\eqn\ex{
\sum_{\sigma \in S_{n} } X_{\sigma} \chi_{R} (\sigma) =
\sum_{R'\in Y_n^*}
 \delta (R,
R') (-1)^{n/2}, }
 the delta function is $1$ if $R$ and $R'$ are the
same and
zero otherwise. From the orthogonality relation,
\eqn\orthog{ \sum_{R\in Y_n} \chi_{\scriptscriptstyle R} (\sigma)
\chi_{\scriptscriptstyle R} (\tau) =
\delta_{T_{\sigma},T_{\tau}} { {n!}\over {T_{\sigma}} }   ,}
we get \eqn\Xsig { X_{\sigma_1} = (-1)^{n/2} \sum_{R \in Y_n^*}
 {{\chi_{\scriptscriptstyle R}
(\sigma_1)} \over {n!} } .}

Now if $\sigma$ is a permutation of $n$ elements and it can be written as
$\sigma_1 \sigma_2$ in cycle notation,  where $\sigma_1 \in S_{n_1}$
$\sigma_2 \in S_{n_2},n_2=(n-n_1)$ i.e $\sigma_1\sigma_2$ also lives
in the subgroup
$S_{n_1}\times S_{n_2} $ of $S_n$, then we can write
 \ref\sagan{B. E. Sagan, The
Symmetric Group: Representations, Combinatorial Algorithms,
 and Symmetric
Functions,' Pacific Grove, California: Wadsworth and Brooks /Cole Advanced
Books and Software, 1991.}\ref\FH{W. Fulton and J. Harris, `Representation
Theory,' Springer-Verlag New York Inc. 1991},
\eqn\rec{\chi_{\scriptscriptstyle R}( \sigma ) = \sum_{R_1\in Y_{n_1},
R_2 \in Y_{n_2} }
 g(R_1,R_2;R) \chi_{\scriptscriptstyle R_1} (\sigma_1)
\chi_{\scriptscriptstyle
R_2} (\sigma_2) }
The existence of such an expansion follows from the fact that
the left hand
side is a class function for the subgroups $S_{n_1}$ and $S_{n_2}$.
The fact
that the coefficients are precisely the Littlewood-Richardson
coefficients
follows from applications of the Frobenius reciprocity theorem \sagan.

  Now we can use equations \charform ,
\ex\ and \rec , to write the $O(N)$ character formula as follows:
\eqn\onrew{ \chi_{\scriptscriptstyle R} (U) = \{R\} +
\sum_{\scriptstyle n_1\ge 2}\sum_{\scriptstyle n_2\ge 0\atop
 \scriptstyle n_1+n_2=n}
\sum_{\sigma_1 \in S_{n_1}, \sigma_2 \in
S_{n_2} }    X_{\sigma_1} {{ Y_{\sigma_2} (U)} \over {(n_2)!} }
  \chi_{\scriptscriptstyle R} (\sigma_1 \sigma_2)        .}
A more compact way of writing
 the formula absorbs the first term by allowing $n_1$ to be $0$ as well,
defining $X_{\sigma_1}=1$ for $n_1=0$, to
give
\eqnn\onchar $$
\eqalignno{   \chi_{\scriptscriptstyle R} (U) &=
\sum_{\scriptstyle n_1\ge 0}\sum_{\scriptstyle n_2\ge 0\atop
 \scriptstyle n_1+n_2=n}
 \sum_{\sigma_1 \in S_{n_1}, \sigma_2 \in
S_{n_2} } X_{\sigma_1}{{Y_{\sigma_2} (U)} \over {(n_2)!} }
  \chi_{\scriptscriptstyle R} (\sigma_1 \sigma_2) \cr
&=
\sum_{\sigma\in S_n} \chi_{\scriptscriptstyle R}(\sigma)
\sum_{{\scriptstyle n_1 \ge 0}}
\sum_{ {\scriptstyle T_{\sigma_1}} \atop
        {\scriptstyle {T_{\sigma_1} T_{\sigma_2} = T_{\sigma}}}
     }
  X_{\sigma_1} {{Y_{\sigma_2} (U)}
\over {(n_2)!} }   {{|T_{\sigma_1}| |T_{\sigma_2}|}\over
{|T_\sigma|} } \cr
&= \sum_{\sigma \in S_{n}} \chi_{\scriptscriptstyle R}(\sigma)
\tilde{Y}_{\sigma}(U)  .&\onchar\cr}$$
In the second line the sum is over all ways of separating
 the cycles of each
$\sigma \in S_n $ into 2 sets of cycles \foot{ I am grateful to
S. G. Naculich, H. A. Riggs, and H. J. Schnitzer for a correspondence
 pointing
out a subtlety in going from the first to the second line of \onchar ,
 which was missed
in the first version of this paper.}
. The first set of cycles
 corresponds
to a conjugacy class in $S_{n_1}$, a representative
 of which is $\sigma_1$
 and the second set to a conjugacy class
in $S_{n-n_1}$, a representative of which is $\sigma_2$.
 The last line is a definition of $\tilde{Y}_{\sigma}(U)$.
 By putting $U=1$ we get the large $N$ expansion for the
dimension of the $O(N)$ representation \dimens.

\newsec{Geometrical interpretation}
 Having a formula for the $\Omega$ point in the form above,
 the same steps
of Gross-Taylor  can be used to write the
2D  Yang-Mills partition function
on an arbitrary manifold $\Sigma$ as a sum over homomorphisms from
$\Pi_1(\Sigma \setminus \{punctures\})$ into the symmetric group of
 permutations on the
sheets of the cover, where there are punctures for the branch points
and the $|2-2G|$ omega points. The same argument as in \GrTa\ can
be used to show that inverse powers of the dimension are
written in terms
of the inverse powers of $\Omega$. The argument relies on
 the fact that the
element $\Omega_n$ of the group algebra of $S_n$ commutes
with $S_n$, which
is also true here because $\Omega_n$ is a sum of elements
of $S_n$ with
coefficients that only depend on the cycle structure
(conjugacy class) of
the element, as is clear from inspection of \dimens.
Negative powers of the omega point can be
dealt with in a $1/N$ expansion as in \GrTa . For arbitrary
 closed non-orientable surfaces, the reality of the
reps of $S_n$ \NRS\  guarantees that the
 full partition function
 can be expressed as a
sum of homomorphisms to $S_n$.

 We recall from \GrTa\ how the $\Omega$
point can be inserted
in the partition function of the theory.
Take the case of sphere,
\eqnn\omegsph $$\eqalignno
{ {\cal Z} &= \sum_{R} (\dim R)^2 e^{-{{\lambda A}\over {2N}}
C_2(R)} \cr &= \sum_{n=0}^{\infty}\sum_{R\in Y_n} {{N^{2n}}
\over {(n!)^2}}
 \chi{(\Omega_n)}^2 e^{-{{\lambda A}\over {2N}}
 (nN + {{n(n-1)\chi_{\scriptscriptstyle R}(T_2)}\over {d_R}}-n)} \cr
&= \sum_{n=0}^{\infty}\sum_{R\in Y_n}\sum_{i\ge 0}
\sum_{p_1\ldots p_i \in T_2}
 {{N^{2n-i}}\over
{(n!)^2}} e^{-{{n\lambda A }\over 2} }e^{{n\lambda A}\over {2N}}
d_R \chi_{\scriptscriptstyle R}
(p_1\ldots p_i \Omega_n^2)
(-1)^i {{(\lambda A)^i} \over {i!}} \cr
&=  \sum_{n=0}^{\infty}\sum_{i\ge 0}\sum_{p_1\ldots p_i\in T_2}
 N^{2n-i} e^{{{-n\lambda A}\over 2}}
e^{{n\lambda A}\over {2N}} (-1)^i {{(\lambda A)^i}\over {i!}}
[{1 \over {n!}}\delta (p_1\ldots p_i \Omega_n^2)] .&\omegsph\cr }$$
In combining the product of the characters into a single character,
we use
the fact that
$\Omega_n$ commutes with $S_n$. The extra factor $e^{n\lambda A/2N}$
compared to $U(N)$ is accounted by infinitesimal crosscaps
being mapped to a point \NRS. We notice also that for either
$O(N)$ or $SO(N)$ (whose lie algebras are isomorphic) the casimir
does not give rise to collapsed handles and tubes as for $SU(N)$.
 Finally
as observed in \NRS\ the above expansion, which is the analog of chiral
 expansion in
the case of unitary groups, suffices to give non trivial
 anwers on both
orientable and non orientable surfaces, unlike the $SU(N)$ case.
 This is
related to the reality of the reps of $O(N)$.

For a manifold with $G$ handles and $b$ boundaries
 the appropriate observables for a stringy
interpretation are
\eqnn\bound $$\eqalignno { Z(G,\lambda A, N;\{\sigma_i\})
&= <\prod_i \tilde{Y}_{\sigma_i}(U_i)>\cr
&= \int dU_1dU_2\ldots dU_b Z(G,\lambda
A;U_1,\ldots, U_b) \prod_i \tilde{Y}_{\sigma_i}(U_i) .&\bound\cr}  $$
The orthogonality property of the $\tilde{Y}_{\sigma}$ will
 follow from that of
$O(N)$ and $S_n$ characters, as shown for the analogous functions
 in the unitary case
 \GrTa . This orthogonality will guarantee \GrTa\ that the partition
function defined above will count maps to the target space with
permutations in the conjugacy class of $\sigma_i$ on
 the sheets covering the boundaries.

We now have to find a geometrical interpretation for all the
coefficients
appearing in $\Omega_n$ of equation \Omegapt . For terms for
which $n_1=0, n_2=n,$
 the coefficient of
$\sigma_2=\sigma \in S_n$ is $N^{K_{\sigma}-n}$. This has a simple
interpretation in terms of multiple branch points, and the power of $N$
correctly gives the Euler character of a surface branched in the
way prescribed by $\sigma$.  The product of ${{T_{\sigma}}\over{(n)!}
}$ (which is equal to ${1\over {n!}}$ times the number of distinct
 homomorphisms describing the same configuration of unlabelled sheets )
with the
 coefficient of $N^{K_{\sigma}-n}\sigma$ in the omega point gives
 ${1\over {|S_{\nu}|}}$  where $|S_{\nu}|$ is the symmetry factor for
the branched covering described by $\sigma$.
 This kind of singularity and symmetry factor  are familiar from the
chiral omega point of $U(N)$ or $SU(N)$ \GrTa .
When $n_1$ is not zero, we have a  sum over decompositions of the
cycles of $\sigma$ into 2 sets, such that the first set can be associated
with a permutation $\sigma_1$ in the class $T_{\sigma_1}$
of $S_{n_1}$ and the second with a permutation $\sigma_2$
in the class $T_{\sigma_2}$ of $S_{n_2}$.
 Now ${1\over {(n)!}}T_{\sigma} $ (which is again ${1\over {n!}}$
 times the number of distinct homomorphisms into $S_n$ describing the same
configuration of unlabelled sheets)  multiplied by
the coefficient of a given decomposition of the cycles of $\sigma$
into two sets, separates into
$|T_{\sigma_1}|X_{\sigma_1}N^{-n_1}$
times ${{|T_{\sigma_2}|N^{K_{\sigma_2}-n_2}}\over {n_2!}}$.
 The latter term
 again correctly describes the genus and symmetry factors
 for multiple branch
points described  by  $\sigma_2$.
The coefficient of $\sigma_1$ has a power of $N$ which is
smaller than the one expected from the branch points responsible for the
permutation. A similar situation is met in the $U(N)$
  case for the coupled
$\Omega$ point where there are collapsed tubes connecting cycles of the
same length. So this is a useful ingredient, and
we call the $n_1$ sheets the singular sector which we will
 now study.

We will need one more singular object. Let us look
at a configuration that can arise from the $O(N)$ omega
point, for concreteness,
say on a target space which has the topology of a disc, and consider
terms where there is no power of the area so that there are no branch
points coming from the casimir. Let us further specialise to a
term where $n=n_2$ and $\sigma_2$ is made of one cycle of length $n_2$.
This term contributes to $\dim R$ a factor
$N^{K_{\sigma_2}}=N^1= N^{2-2g-b}$, where $g$ and $b$
are the number of handles and boundaries, respectively, of
the worldsheet.
If we specialised instead to a term where $n=n_1$ and $\sigma_1$
is made of
one cycle only, then the associated power of $N$ is zero. This can be
understood if the branch point comes with a tiny crosscap as well.
We also observe that $n_1$ is even,
so odd total number of sheets do not appear in the singular sector (which
we recall are the sheets on which $\sigma_1$ are acting).
 This suggests that these
crosscaps can only live on even cycles. The power of $N$ associated with
$\sigma_1$ is always correctly accounted for, if all the
 cycles either have a
crosscap or share a tube with another cycle.
 We will consider the implications
of this simplest set of singularities needed to account for the powers of
$N$, for example we do not invoke pants connecting more than two cycles or
such higher contact terms.
And we will show that the few
singularities can correctly account for the
$X_{\sigma_1}$. Now what kind of tubes are allowed? Do they only connect
cycles of the same length as in the coupled $U(N)$ case ? We will
 look at some
simple examples of the coefficient $X_{\sigma_1}$ to understand the
symmetry factors associated with each singularity and then use the
geometrical picture that emerges to guess a general formula for
$X_{\sigma_1}$. We will prove this formula in the next section.

 We now show that for permutations $\sigma_1$, made of one or two cycles,
the quantity
$|T_{\sigma_1}| X_{\sigma_1} \equiv TX_{\sigma_1}$ ($|T_{\sigma_1}|$
is the number of elements in the conjugacy class of $\sigma_1$)
 is the symmetry factor
for some singular worldsheet configurations involving multiple branch
points, crosscaps, and tiny tubes connecting two cycles.
We can use the sum of characters given in \Xsig\ when
 $\sigma_1$ is a permutation
made of one cycle only, or two cycles, appendix A describes some details.
 For one cycle of even length $2m$,
$TX_{\sigma_1} = {{-1}\over {2m}}$. So each crosscap in an $\Omega$
point comes with a
minus sign. Note that the `crosscap with branch point' in the
$\Omega$ point comes with the same symmetry factor as a
 pure branch point in the
$\Omega$ point, but has an extra minus sign.
 For two odd cycles of different
lengths the $TX$ vanishes, which is consistent
with the idea that odd cycles with crosscaps do not
occur and that tubes only connect cycles of equal length.
 For two even cycles of unequal length it is just the product of the
symmetry factors for each, suggesting again that cycles
 of different lengths do
not interact. For two cycles of equal odd length $r$,
$TX= {{-1}\over {2r }}$. For two cycles of equal even length $r$,
$TX= {{-1} \over {2r}} + {1\over{2r^2} }$. The first term in each case
comes from a collapsed tube connecting the $2$ cycles. Except for the
factor of $2$, the symmetry factor and the
minus sign are exactly the same as for a
tube connecting two
sheets of opposite orientation in the coupled $\Omega$
point of $U(N)$. The extra symmetry is expected because the sheets
being connected are not equipped with different orientations as
in the case
of $U(N)$. The additional term in the case of two even cycles comes from
configurations where the two even cycles are each
carrying a crosscap.
 Note
that we do not have separate factors for infinitesimal Klein bottles and
 tori because in
the singular limit they are the same.

We now summarise the complete set of singularities in the $O(N)$ omega
point.
In the singular sector (described by $\sigma_1$)
 then, an even cycle has to carry a crosscap or be
connected to another cycle of the same length by a collapsed tube. An odd
cycle must be connected to another cycle of the same length. Each crosscap
or collapsed tube comes with a minus sign. These singular configurations
come with the natural symmetry factors. The sheets not in the singular
sector (described by $\sigma_2$ ) come with multiple branch points. The
only new singularity compared to the coupled
 $U(N)$ case is the `crosscap on even
cycles'. These singular configurations are shown in figure 1.

This predicts that the $TX_{\sigma}$ should factorise into separate
factors for each cycle length. And the contribution is zero from an odd
number of odd cycles. For an even number $2m$  of odd cycles of length
$r$, it should be
\eqn\symodd{ TX_{[(r^{2m})]}= (-1)^m { 1 \over{(m)!2^m r^m} } .}
The minus signs come from the $m$ tubes, $1/(2r)$ is the symmetry factor
of a pair of cycles connected by a tiny tube, and $m!$ comes from the
symmetry operation of permuting the $m$ identical pairs of cycles.
For  any number $n$ of even cycles  (order $r$), we sum over
 $i$ collapsed handles (going
from zero to the integer part of $(n/2)$), with
$(n-2i)$ crosscaps, to get
\eqnn\symeven $$\eqalignno
 { TX_{[r^n]} &=  \sum_{i=0}^{int (n/2)} {
{(-1)^{n-i}}\over{i!(n-2i)!2^ir^{n-i} } }\cr
&= \sum_{i=0}^{int (n/2)}
{{(-1)^{n-2i}} \over {(n-2i)! r^{n-2i}} } \times
{ {(-1)^i}\over{i!r^i2^i} }
,&\symeven\cr}$$ where, in the last line, we have separated out
 the factors for crosscaps and tubes.

\newsec{Derivation of Formulae for the Sum of Characters $X_{\sigma}$ }
Directly summing the characters to get \symodd\ and \symeven\
in the case of arbitrary permutations
$\sigma_1$ appears hard.
The following identity between symmetric functions \lit\ will be very
useful:
\eqnn\ident $$\eqalignno{ \prod_{i <j} (1-\alpha_i\alpha_j)
\prod_{i} (1-\alpha_i^2) &=
1 + \sum_{R \in Y^*} (-1)^{n/2} \{R\} \cr &= 1 + \sum_{n \ge 2}
\sum_{R\in Y_n^*}   (-1)^{n/2}   \sum_{\sigma \in S_n}
{ {\chi_{\scriptscriptstyle R}(\sigma)} \over
{n!} }P_{\sigma}(\alpha_1,\alpha_2\cdots)  \cr
     &= 1 + \sum_{n\ge 2}\sum_{\sigma \in S_n} X_{\sigma} P_{\sigma}
.&\ident\cr}$$
We first determine $X$ for $ \sigma_1 $  made of $n_1$
 cycles all of length
$r_1$. To start we consider the case $r_1=1$. We only need to keep
$n_1$ distinct $\alpha $'s in the identity \ident , and
 we need to find the coefficient of the power sum symmetric
polynomial $(\alpha_1+\alpha_2+\cdots+\alpha_{n_1})^{n_1}$.
 This polynomial is the only one containing the monomial
$\alpha_1\alpha_2\cdots\alpha_{n_1}$. On LHS the
monomial can be built by taking pairs of $\alpha$'s from the different
factors. The number of ways of doing that is equal to the number of ways of
pairing up all the elements from $n_1$ objects which is zero if $n_1$ is
odd, and if $n_1=2m_1$ we get the identity.

\eqnn\cycone $$\eqalignno{ (-1)^{m_1} {{(2m_1)!}\over{2^{m_1} (m_1)!}}
&= (2m_1)!
TX_{[(1^{2m_1})]}\cr \Rightarrow TX_{[(1^{2m_1})]} &= {{(-1)^{m_1}}\over
{2^{m_1}m_1!}}.&\cycone\cr}$$ This agrees with \symodd . Figure 2
 illustrates
the case $m_1=5$ and compares with a similar configuration in the $U(N)$
omega point.
For general $r_1$ we choose the $r_1n_1$ indeterminates in the
 following way
\eqn\vars{\alpha_{i+kn_1} = \alpha_{i}e^{2\pi i k/r_1}}
 where $i$ runs from $1$ to
$n_1$; and $k$ runs from $0$ to $r_1-1$. This guarantees
that $(\alpha_1^K  +
\alpha_2^K+\cdots + \alpha_{r_1n_1}^K)$ vanishes unless $K$ is
a multiple of
$r_1$. This means that the only power sum symmetric polynomial
which contains the monomial $\alpha_1^{r_1}\alpha_2^{r_1}\cdots
\alpha_{n_1}^{r_1}$ is $(r_1\alpha_1^{r_1}+r_1\alpha_2^{r_1}+\cdots +
r_1\alpha_{n_1}^{r_1})^{n_1}$ whose coefficient is $TX_{[(r_1)^{n_1}]}$.
With our choice of variables, the LHS can be
written, for odd $r_1$,
as
 \eqn\Lcyc{ \prod_{1 \le i<j\le n_1} (1-\alpha_i^{r_1}
\alpha_j^{r_1}) ^{r_1}
 \prod_{1 \le i\le n_1} ( 1 - \alpha_i^{2r_1})^{(r_1+1)/2} ,}
and, for even $r_1$, as :
\eqn\Lecyc{ \prod_{1 \le i<j\le n_1} (1-\alpha_i^{r_1}\alpha_j^{r_1})^{r_1}
 \prod_{1 \le i\le n_1} ( 1 - {\alpha_i}^{2r_1})^{r_1/2} (1-{\alpha_i}
^{r_1}).}
Comparing coefficients of the monomial gives the
answer in \symodd { }and \symeven .

Now we consider the proof of factorisation. Consider for a start the
case of $n_1$ cycles of length $r_1$ and $n_2$ cycles of length $r_2$.
Choose the variables $$\alpha_1\cdots \alpha_{r_1n_1+r_2n_2}$$ as follows.
Let \eqna\varsto
$$\eqalignno{ \alpha_{i+k_1n_1} &= \tilde{\alpha_{i}}
                              e^{{2\pi ik_1} \over {r_1}},\cr
                            &1 \le i \le n_1,\cr
                            & 0 \le k_1 \le (r_1-1),&\varsto a\cr
 \alpha_{i+r_1n_1+k_2n_2} &= \tilde{\beta_{i}}e^{{2\pi ik_2} \over
{r_2}},\cr
                            &1 \le i \le n_2, \cr
                            & 0 \le k_2 \le (r_2-1)&\varsto b.\cr }$$
With this choice there is only one term on the RHS containing the monomial
\eqn\monom
{\tilde{\alpha_1}^{r_1}\tilde{\alpha_2}^{r_1}\cdots
\tilde{\alpha_{n_1}}^{r_1}
\tilde{\beta_1}^{r_2}\tilde{\beta_2}^{r_2}\cdots
\tilde{\beta_{n_2}}^{r_2}
.}
And it is the polynomial $$
( \alpha_1^{r_1} +  \alpha_2^{r_1} \
+ \cdots  \alpha_{r_1n_1+r_2n_2}^{r_1})^{n_1}
( \alpha_1^{r_2} +  \alpha_2^{r_2} +\cdots
\alpha_{r_1n_1 + r_2n_2}^{r_2})^{n_2},$$ whose coefficient is
$TX[(r_1^{n_1},r_2^{n_2})]$.
Now suppose without loss of generality that $r_2>r_1$. Then powers of
$\tilde{\beta}$
in the monomial we are looking for cannot come from the first $n_1$
factors, so must be chosen from the last $n_2$ factors only. So the powers
of $\tilde{\alpha}$ have to come from the first $n_1$. The combinatoric
factor on the RHS is thus the
product of those that determined the $TX[(r_1^{n_1})]$ and
$TX[(r_2^{n_2})]$.  Now on the LHS of \ident\ we can separate out
the product into terms
containing $\tilde{\alpha}$ only and terms containing $\tilde{\beta}$ only,
and mixed terms.  The mixed terms can be computed to be
$$\prod_{1\le i\le n_1,1\le j\le n_2}
(1-\tilde {\alpha_i}^{ {{r_1r_2}\over r}}
\tilde {\beta_j}^{{r_1r_2}\over r} )^r,$$ where $r$ is the greatest common
divisor of $r_1$ and $r_2$.    This does not contribute to the monomial
\monom . This argument for factorisation clearly generalises to the
the case of an arbitrary number of cycle lengths.   We have proved then,
that there are no infinitesimal tubes  connecting cycles of different
lengths.

For spinor representations \lit, the dimension can be
 written as $2^N$ times an
expression of the form  \dimens\ with $X_{\sigma_1}$
replaced by $\hat X_{\sigma_1}$ whose
generating function is the expression \lit,
\eqn\spin{ \prod_{i} (1-\alpha_i) \prod_{i<j} (1-\alpha_i \alpha_j).}
The same steps as above shows that similar expressions for
 $\hat X_{\sigma_1}$ can be written as a sum over configurations. The only
difference compared to the case of tensor representations
 is that the crosscaps live on odd
cycles only.

\newsec{ About $Sp(N)$ }
Many similarities between $O(N)$ and $Sp(N)$ have been observed in the
context of matrix models \ref\Brez{E. Brezin and H. Neuberger,
`Multicritical Points of Unoriented Random Surfaces,' LPS-ENS-245,
RU-90-39, Nuc. Phys. B350 513-553, 1991.}
 and loop equations \ref\Migtwo{A. A. Migdal,
`QCD=Fermi String Theory,' Nuclear Physics B189 (1981) 253-294.}.
One interesting
relation between the dimensions of $O(N)$ and $Sp(N)$
tensor representations will be useful here in giving
a quick answer for the $Sp(N)$ case using
the result for $O(N)$. From \ref\dunne{G. V.
Dunne, `Negative dimensional groups in quantum physics,' J. Phys. A : Math.
Gen. 22 (1989) 1719-1736.}
\ref\CK{P. Cvitanovic and A. D. Kennedy, `Spinors in
Negative Dimensions,' Phys. Scr. 26, (1982) 5.}
\ref\king{R. C. King, `The dimensions of
irreducible tensor representations of the orthogonal and symplectic groups,'
Can. J. Math., Vol. XXIII, No.1, 1971, pp. 176-188. } we have
\eqn\spn{ \dim_{[Sp(N)]} R = (-1)^n\dim_{O(-N)} \tilde R ,}
 where $\tilde R$ is the conjugate representation,
 related to $R$ by exchanging rows
with columns, and $\dim_{O(-N)}$ is meant in the sense of analytic
continuation. Using this equation and \dimens\ we can write
 the following equation
for
the dimension of a representation of $Sp(N)$
\eqnn\dimensp $$\eqalignno{
\dim R &= \sum_{\sigma \in S_{n}}
{ {\chi_{\scriptscriptstyle \tilde R}(\sigma)}\over {n!} }
\sum_{ {\scriptstyle n_1 \ge 0} }
\sum_{ {\scriptstyle T_{\sigma_1}} \atop
        {\scriptstyle {T_{\sigma_1} T_{\sigma_2} = T_{\sigma}}}
     }
(-N)^{K_{\sigma_2}} X_{\sigma_1} {{n!} \over {n_2!}} (-1)^n
{{|T_{\sigma_1}| |T_{\sigma_2}|}\over {|T_\sigma|}}\cr
 &= \sum_{\sigma \in S_{n}}
{ {\chi_{\scriptscriptstyle R}(\sigma)}\over {n!} }
\sum_{ {\scriptstyle n_1 \ge 0} }
\sum_{ {\scriptstyle T_{\sigma_1}} \atop
        {\scriptstyle {T_{\sigma_1} T_{\sigma_2} = T_{\sigma}}}
     }
 (-1)^{ K_{\sigma_1} }
      N^{K_{\sigma_2}}
 X_{\sigma_1} { {n!} \over {n_2!} }
{{|T_{\sigma_1}| |T_{\sigma_2}|}\over {|T_\sigma|}}.
&\dimensp\cr}$$
We used the fact \lit\ that $\chi_{\scriptscriptstyle \tilde R} (\sigma)=
\chi_{\scriptscriptstyle R} (\sigma)(-1)^p$ where $p$ is $0$ if the
permutation is even and $1$ if it is odd,
and we wrote
$(-1)^p=(-1)^{n-K_{\sigma}}=(-1)^{n-K_{\sigma_1}-K_{\sigma_2}}$.
This means that the geometrical
interpretation of the $\Omega$ point of $Sp(N)$ is identical except for
a minus sign $(-1)^{K_{\sigma_1}}$. From the geometrical interpretation
of $X_{\sigma_1}$,
this is equivalent to saying that
the collapsed crosscaps associated with branch points do not come with a
minus sign.
 The natural generalisation of \dimensp\ to arbitrary characters is,
\eqn\spnchar{
\chi_{\scriptscriptstyle R}(U) = \sum_{\sigma \in S_{n}}
{ {\chi_{\scriptscriptstyle R}(\sigma)}\over {n!} }
\sum_{ {\scriptstyle n_1 \ge 0} }
\sum_{ {\scriptstyle T_{\sigma_1}} \atop
        {\scriptstyle {T_{\sigma_1} T_{\sigma_2} = T_{\sigma}}}
     }
(-1)^{K_{\sigma_1}} X_{\sigma_1} Y_{\sigma_2}(U){{n!} \over
{n_2!} }{{|T_{\sigma_1}| |T_{\sigma_2}|}\over {|T_\sigma|}} ,}
where $U$ is now in $Sp(N)$. We need this equation for a geometrical
interpretation of the theory on a manifold
with boundaries and it can be proved
by using the same steps as in the $O(N)$
case, as we now outline. It is pointed
 out in \king\ that the analog for $Sp(N)$ of \charform\
differs only in that the set $Y^*$ is replaced by its conjugate, which is
the
set that, in Frobenius notation (see appendix A), consists of Young tableaux
of the form
\eqn\dual{ \pmatrix { a & b & c &\ldots \cr a+1 & b+1 & c+1 &\ldots }.}
We can repeat the steps in section 2 to write a formula for
$X_{\sigma_1}(Sp(N))\equiv \tilde X_{\sigma_1}$
 for which the generating function replacing the one
on the LHS of \ident\ is $\prod_{i<j} (1-\alpha_i\alpha_j)$\lit.
 By the same
choice of variables as in section 4 we can derive the analog of equation
\Lcyc\ which is
\eqn\SpLcyc{ \prod_{1 \le i<j\le n_1} (1-\alpha_i^{r_1}
\alpha_j^{r_1}) ^{r_1}
 \prod_{1 \le i\le n_1} ( 1 - \alpha_i^{2r_1})^{(r_1-1)/2} ,}
and the analog of \Lecyc\ which is
\eqn\SpLecyc{
\prod_{1 \le i<j\le n_1} (1-\alpha_i^{r_1}\alpha_j^{r_1})^{r_1}
 \prod_{1 \le i\le n_1} ( 1 - {\alpha_i}^{2r_1})^{{{r_1}\over 2}-1}
(1+{\alpha_i}
^{r_1}).}
We observe that the terms which contribute to the monomial
$\alpha_1^{r_1}\alpha_2^{r_1}\cdots
\alpha_{n_1}^{r_1}$ are unchanged except that the terms
which correspond to
crosscaps sitting on branch points have a
plus sign instead of a minus sign.
The proof of factorisation is identical.
Note that
crosscaps coming from the casimir are also of opposite
sign for $O(N)$
and $Sp(N)$ \NRS .

\newsec{Conclusions and Speculations}
In performing the large $N$ expansion of the partition function for the
case of $U(N)$ gauge group, the naive expansion
analogous to \omegsph\ did not correctly reproduce the large $N$
asymptotics. One of its most obvious weaknesses \GrTa\ was that it gave
a trivial answer for nonorientable target spaces
whereas there is no
reason to expect the large $N$ asymptotics of the formulae
derived in \ref\wit{E. Witten, `On Quantum Gauge Theories In
Two Dimensions,' Commun. Math. Phys. 141, 153 (1991).}\ to be
trivial. The coupled $\Omega$ point \GrTa\ is necessary
in order to get the correct large $N$, large $A$, approximation.
 For $O(N)$ and $Sp(N)$ the naive expansion does not give
trivial answers \NRS\  on nonorientable target spaces.
Moreover we have found
close similarities
between the structure of the $O(N)$  and $Sp(N)$ $\Omega$ points  with the
coupled $U(N)$ $\Omega$ point which suggests that,
the `naive'
expansion might give the correct large $N$, large $A$
asymptotics for the $O(N)$ and $Sp(N)$ gauge groups.
The methods of \ref\DK{M. Douglas and V. Kazakov, `Large N Phase
Transition in Continuum $QCD_2$,' LPTENS-93/20, RU-93-17.}, as applied to
$O(N)$ or $Sp(N)$ could perhaps be used to see if the large $N$ large $A$
asymptotics for these gauge groups is indeed correctly given by the
expansion in \omegsph . For small $A$ the spinor representations with the
factor of $2^N$ in their dimension formula will probably have to be taken
into account.

Our results on the $\Omega$ point together with those of \NRS, suggest
 that the string action of QCD2 must be of a
form that is generalisable to describe $O(N)$ and $Sp(N)$
 Yang Mills. Whereas many questions about
2D Yang Mills  are most easily answered by
starting with the original Yang Mills action, it seems that a satisfactory
answer to the question of why there are no higher order contact terms than
crosscaps and tubes connecting cycles of equal length can most naturally
come from a string picture for all these theories.

One interesting
fact about the infinitesimal tubes is that they are not of two distinct
kinds, Klein bottle and torus type (although the theory does contain
worldsheets with both topologies \NRS). This is naturally
understood if the Omega point is associated with
 singular objects on the worldsheet rather
than just being a singularity of the map from worldsheet to target space.
A degenerated torus is indistinguishable
from a degenerated Klein bottle, whereas singular maps from a Klein bottle
to a point and a torus to a point are distinct. This suggests that the
ingredients that go into the omega point, infinitesimal crosscaps and
infinitesimal tubes, are closely connected to the local worldsheet physics,
and perhaps the local operator content of
the worldsheet theory. Further tests of this picture should be made,
perhaps in the framework set up by Kostov \ref\Kostov{I. Kostov,
`Continuum $QCD_2$ in terms of Discretized Random Surfaces with Local
Weights,' SPhT/93-050.}.

Identifying the appropriate observables for manifolds with boundary
may be of help in developing a Das-Jevicki like picture for the $O(N)$ and
$Sp(N)$ case, as done for $U(N)$ in \ref\MinPol{J. A. Minahan and A.
 Polychronakos, `Equivalence of Two Dimensional QCD and the $c=1$ Matrix
Model,' UVA-HET-93-02.}
\ref\Doug{M. R. Douglas, `Conformal Field Theory techniques for
Large N Group Theory,' RU-93-13, NSF-ITP-93-39. }. It should also allow a
computation of Wilson loops by the gluing  method of \GrTa, adapted to
$O(N)$ and $Sp(N)$.

\bigbreak\bigskip\bigskip\centerline{{\bf Acknowledgements}}\nobreak
I would like to thank Professor Gregory Moore for many discussions and for
many comments on the preliminary drafts. I would like to thank Jim Horne
and Ronen Plesser for many discussions. This work was supported by the DOE
grant DE-AC02-76ER03075.

\appendix{A} {Determination of $X_{[\sigma_1]}$ by Summing Characters}
At this point we need to describe
in more detail the set $Y^*$, which is done conveniently using the
Frobenius notation. The Frobenius notation for Young
tableaux describes them by an array of pairs of numbers. The number of
pairs is equal to the number of boxes on the leading diagonal of
the tableaux.
The upper number of the pair is the number of boxes to the right
 of that box
and the lower number is the number of boxes below that box. A tableau with
row lengths $[6,4,3]$, for example, is described by
$$\pmatrix{5&2&0\cr2 &1&0}.$$ The set $Y^*$ consists of tableaux which
are of the form $$\pmatrix{a+1 & b+1 & c+1 & \ldots \cr a & b & c &
\ldots}.$$ Clearly they can only have an even number of boxes. Using a
formula for characters of a single cycle \sagan , only one tableau in
$Y^*$, with a single element along the diagonal,
will contribute to $X_{\sigma}$ when $\sigma = (2m)$. We find that
$X[(2m)]= { {-1} \over {(2m)!} }$. The sum over all elements in this
conjugacy class gives   ${ {-1} \over {(2m)}}$.

  For conjugacy classes with two cycles $(n_1,n_2), n_1\ne n_2 $
 the sum over characters can still
be done fairly easily , using for example the Murnaghan-Nakayama
recursion
formula for characters of $S_n$ \sagan. Only tableaux with at most two
boxes on the leading diagonal contribute. And we find
 $$ TX[(n_1,n_2)]= { {1}\over {n_1n_2}}$$
if they are both even, and zero if they are both
odd.

\listrefs

\end